\documentclass[aps,prl,twocolumn,showpacs,amsmath,amssymb,superscriptaddress,floatfix]{revtex4}
\usepackage{graphicx,epsf}
\usepackage{bm}

\begin{document}

\title{Inner and outer edge states in graphene rings: A numerical investigation}

\author{D. A. Bahamon}
\author{A. L. C. Pereira}
\author{P. A. Schulz}
\address{Instituto de F\'\i sica, Universidade Estadual de Campinas - UNICAMP,  C.P. 6165, 13083-970, Campinas, Brazil}

\date{\today}

\begin{abstract}

We numerically investigate quantum rings in graphene and find that their electronic properties may be strongly influenced by the geometry, the edge symmetries and the structure of the corners. Energy spectra are calculated for different geometries (triangular, hexagonal and rhombus-shaped graphene rings) and  edge terminations (zigzag, armchair, as well as the disordered edge of a round geometry). The states localized at the inner edges of the graphene rings describe different evolution as a function of magnetic field when compared to those localized at the outer edges.  We show that these different evolutions are the reason for the formation of sub-bands of edge states energy levels, separated by gaps (anticrossings).
It is evident from mapping the charge densities that the anticrossings occur due to the coupling between inner and outer edge states.

\end{abstract}



\maketitle


\section{I. Introduction}

Graphene is a bona fide two-dimensional material showing a great versatility due to its unconventional electronic properties and promising applications to nanoelectronics \cite{geim_rev,rev_mod_phys}. Among the promises is the possibility of structuring graphene at a mesoscopic length. Indeed, some groups have already demonstrated that graphene can be cut in many different shapes and sizes, opening the door to the fabrication of graphene nano-devices through the impressive experimental obtention of graphene quantum dots \cite{novoselov,ponomarenko,ensslin}, quantum rings \cite{morpurgo} and even antidot arrays \cite{schen}. This perspective leads to interesting scenarios since the electronic properties of graphene are deeply influenced by its size and shape: as it is well known for over 10 years, graphene nanoribbons have different properties depending on their width or on their edge terminations \cite{Nakada,Wakabayashi,Brey1,cresti}.
Besides the nanoribbons, some theoretical works have also addressed the effects of confinement on the  electronic structure of graphene quantum dots (flakes) with different geometries, sizes and types of edge \cite{Yamamoto,peeters,akola}.
Recently, the energy levels of graphene quantum rings \cite{recher,abergel} and of graphene antidot lattices \cite{pedersen1,pedersen2} have also been theoretically investigated.

In this paper we numerically analyze the energy spectra as a function of magnetic field ($B$) of graphene quantum dots and graphene quantum rings, with focus on the complex evolution of edge states in the graphene rings.
Here we explore the effects of the interplay among different degrees of freedom given by size, geometry and edge symmetries on the electronic properties of these graphene nanostructures. We consider quantum dots and rings with six (hexagons), three (triangles) and two-fold (rhombus-shaped) rotational symmetry, with  zigzag or armchair edges.  Afterwards we consider round dots and rings, whose edges are not so simply: they are cut in a way to approach the circular geometry. Our attention is concentrated on the continuum limit \cite{ana} of the energy spectra as a function of the magnetic flux of these structures. Edge states appear with energies between consecutive Landau levels (LLs) in such spectra, as could be initially expected \cite{sivan}. However, the interplay between two different edges showing distinct local structures (the quantum rings can be seen as graphene structures containing an antidot, which introduces an inner edge to the system) leads to some surprising subtleties. We observe that the presence of the antidot introduces additional edge states, with a different evolution under $B$: their energies are increased with increasing $B$. For a better understanding of this behavior, the electronic densities of these states are mapped, and we show that edge states that rise in energy with $B$ are located in the internal edges of the ring structure. In this way, we show that inner and outer edge states give origin to the formation of sub-bands separated by energy gaps in the region of the spectra between LLs. The  anticrossing of levels, which defines the sub-bands, occurs due to the inter-edge coupling of states. The formation of sub-bands is highly influenced by symmetry properties, and also by size effects, i.e., the relation between the ring width and the magnetic length.

As will be seen throughout the paper, the choice of quantum rings is strategic since the states within the edge states sub-bands can be perfectly associated to either inner or outer edges, or to a coupling of both edges of the ring structure, therefore enabling a good framework for studying the influence of the edges and edge junctions on the electronic structure and charge distribution.

\section{II. Model}

We use a tight-binding model for a finite two-dimensional honeycomb lattice, considering nearest neighbors hoppings \cite{ana,ana_valley}.  The following non-interacting Hamiltonian is considered:

\begin{equation}
H = \sum_{i} \varepsilon_{i} c_{i}^{\dagger} c_{i}
+ t  \sum_{<i,j>} (e^{i\phi_{ij}} c_{i}^{\dagger} c_{j} + e^{-i\phi_{ij}}
c_{j}^{\dagger} c_{i})
\end{equation}

\hspace{-\parindent}where $c_{i}$ is the fermionic operator on site $i$.
The perpendicular applied magnetic field is included by means of Peierls substitution, which means a complex phase in the hopping parameter ($t$=2.7 eV): $\phi_{ij}= 2\pi(e/h) \int_{j}^{i} \mathbf{A} \! \cdot \! d \mathbf{l} \;$. In the Landau gauge, where the electromagnetic vector potential is defined as $\mathbf{A}=(0,Bx,0)$, one obtains $\phi_{ij}\!=\!0$ along the $x$ direction and $\phi_{ij}\!=\pm \pi (x/a) \Phi / \Phi_{0}$
along the $\mp y$ direction. The magnetic flux ($\Phi$)  per magnetic flux quantum ($\Phi_{0}=h/e$) is defined as: $\Phi / \Phi_{0}=Ba^{2}\sqrt{3}e/(2h)$, and we use $a$=2.46{\AA} as the lattice constant for graphene. The on-site energies are taken as $\varepsilon_{i}=0$.

To consider the ring geometries, a central region of absent atoms (antidot) is defined in the structure by setting the hopping parameters to zero for the absent atoms and the on-site energies at the position of these atoms equal to a large value outside the energy range of the spectra.
The magnetic field we consider is not limited to the central region of the ring, but is homogeneously applied to the entire structure. By exact numerical diagonalization of the Hamiltonian, the energy spectrum as a function of magnetic flux is calculated for different geometries of quantum rings.

\section{III. Edge States of Graphene Rings}

\subsection{A. Antidot effects on the energy spectrum}

We start by calculating the energy spectrum of the structure shown in Fig. 1(a): a finite hexagonal lattice forming an equilateral triangle with zigzag edges. Calling $N_{out}$ the number of individual hexagonal plaquetes along each side of the triangle, the total number of carbon atoms in this structure is $N^{2}_{out} + 4N_{out}+1$ \cite{Yamamoto}.  The energy spectrum for such a triangular graphene quantum dot with $N_{out}$=45 is plotted in Fig. 1(c) as a function of magnetic flux. One can clearly observe the formation of the low energy LLs: the $n$=0 LL at zero-energy, and the $n$=+1 and $n$=+2, with their square-root dependence on magnetic field, typical from graphene systems (the spectrum is symmetrical with respect to the zero-energy).  Also, one can see the expected presence of edge states between consecutive LLs and observe the evolution of these edge states with magnetic flux until coalescing to the LLs \cite{sivan}. The side length of the triangular structure is simply given by $aN_{out}$, where $a$=2.46{\AA} is the lattice constant. So, for the case considered here of $N_{out}$=45, the side length of the triangular dot is $\approx$11nm.

\begin{figure}[t]
\vspace{-0.2cm}
\centerline{\includegraphics[width=8.5cm]{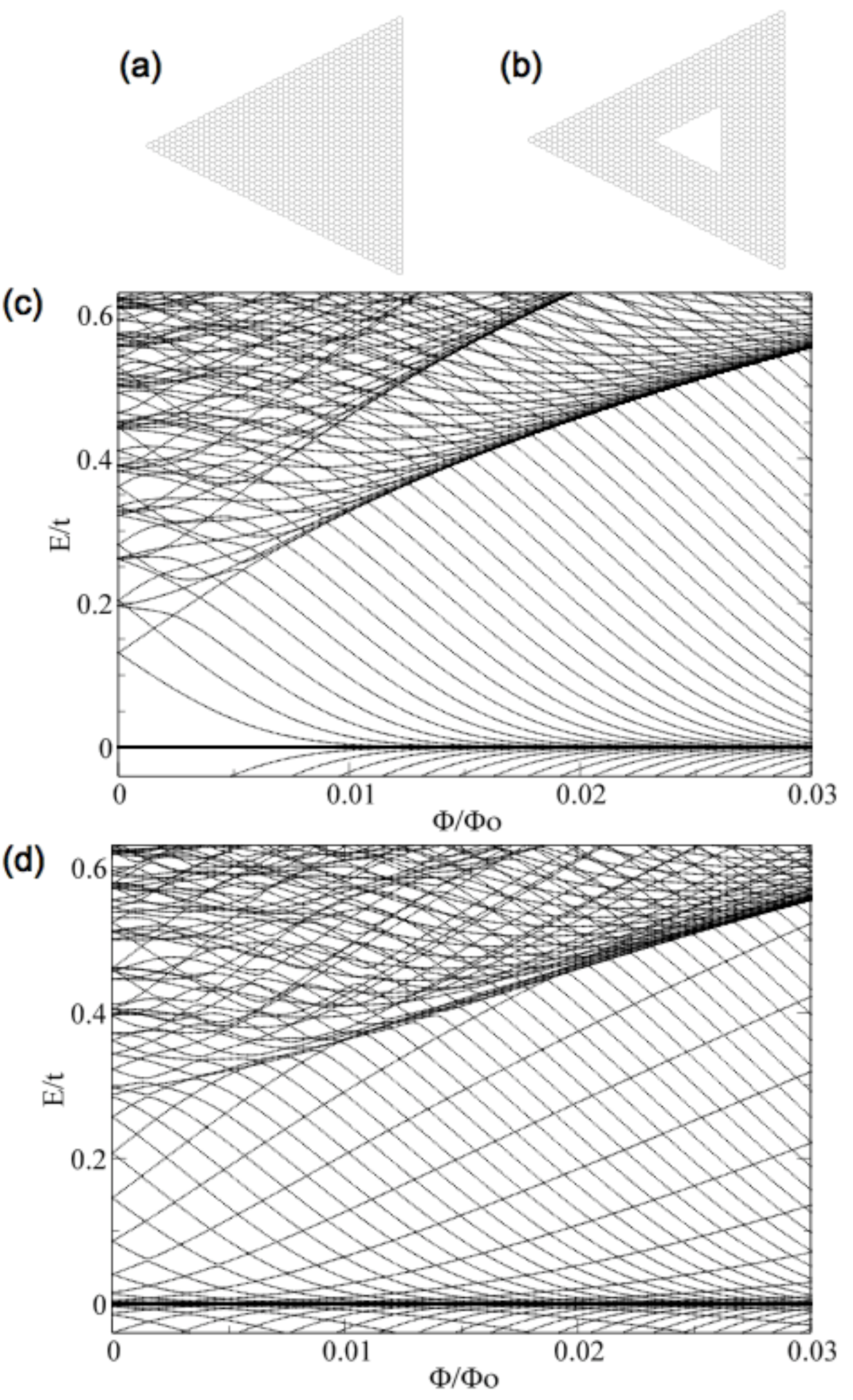}}
\vspace{-0.2cm}
\caption{{\bf (a)} Triangular graphene quantum dot with zigzag edges, with $N_{out}$=45. {\bf (b)} Triangular graphene quantum ring, with $N_{out}$=45 and $N_{in}$=12. {\bf (c)} Energy spectrum as function of the magnetic flux for the structure in (a). {\bf (d)} Energy spectrum as function of the magnetic flux for the structure in (b) }
\end{figure}


We then take this nanostructure as a starting point to develop a triangular  quantum ring just piercing a triangular hole (antidot) in the middle of it. The inner edges of this ring are also zigzag. To define the size of the antidot, we call  $N_{in}$ the number of hexagons at each side of the internal removed triangle. The total number of atoms in this quantum ring is now $N^{2}_{out} + 4(N_{out} - N_{in})-N^{2}_{in} +6N_{in}$. In Fig. 1(b) there is a representation of such a ring for $N_{out}$=45 and $N_{in}$=12, and the corresponding energy spectrum is shown in Fig. 1(d).  The interesting observation is that the presence of the antidot gives origin to additional edge states with a different evolution with magnetic field: states that go up in energy as the magnetic flux is increased. Indeed, looking at vacancies in graphene \cite{ana}, the localized states around such defects also rises in energy with increasing magnetic field, since vacancies are actually minimal antidots. It is also clear that with the introduction of the antidot, the formation of the $n>$0 LLs starts at higher magnetic fields when compared to Fig. 1(c), due to the involved interplay between the inner and outer edge states. On the other hand, this interplay seems to anticipate the formation of the central LL, in these peculiar zero magnetic field limit showing edge states at the Dirac point due to the zigzag structure of the edge \cite{Wakabayashi}.

To analyze in more details how the energy levels of the edge states evolve with magnetic field, in Fig. 2(a) we zoom in the energy scale of Fig. 1(d). It now becomes evident the formation of edge states sub-bands, separated by energy gaps that get smaller with increasing field. One can also note that each of these sub-bands  contains three crossing energy levels for this triangular graphene ring.

\subsection{B. Different evolutions for inner and outer edge states}

In order to gain a deeper understanding of this quite complex evolution of edge states, in Figs. 2(b-d) we look to the wave functions amplitudes of specific edge states. The arrow (b) in the spectrum in Fig. 2(a) is pointing to an edge state whose energy is reduced with increasing $B$. This state is mapped in Fig. 2(b), and clearly is localized at the outer edge of the triangular ring.  The radii of the circles plotted are directly proportional to the wave function amplitude on each site, and we can observe a symmetrical and quasi homogeneous distribution over the edges, with higher concentrations at the outer most lattice sites (and always on the same sublattice, in this case of zigzag edges). A high charge density accumulation is also observed close to each corner forming part of a second charge density belt.


\begin{figure}[t]
\includegraphics[width=7.8cm]{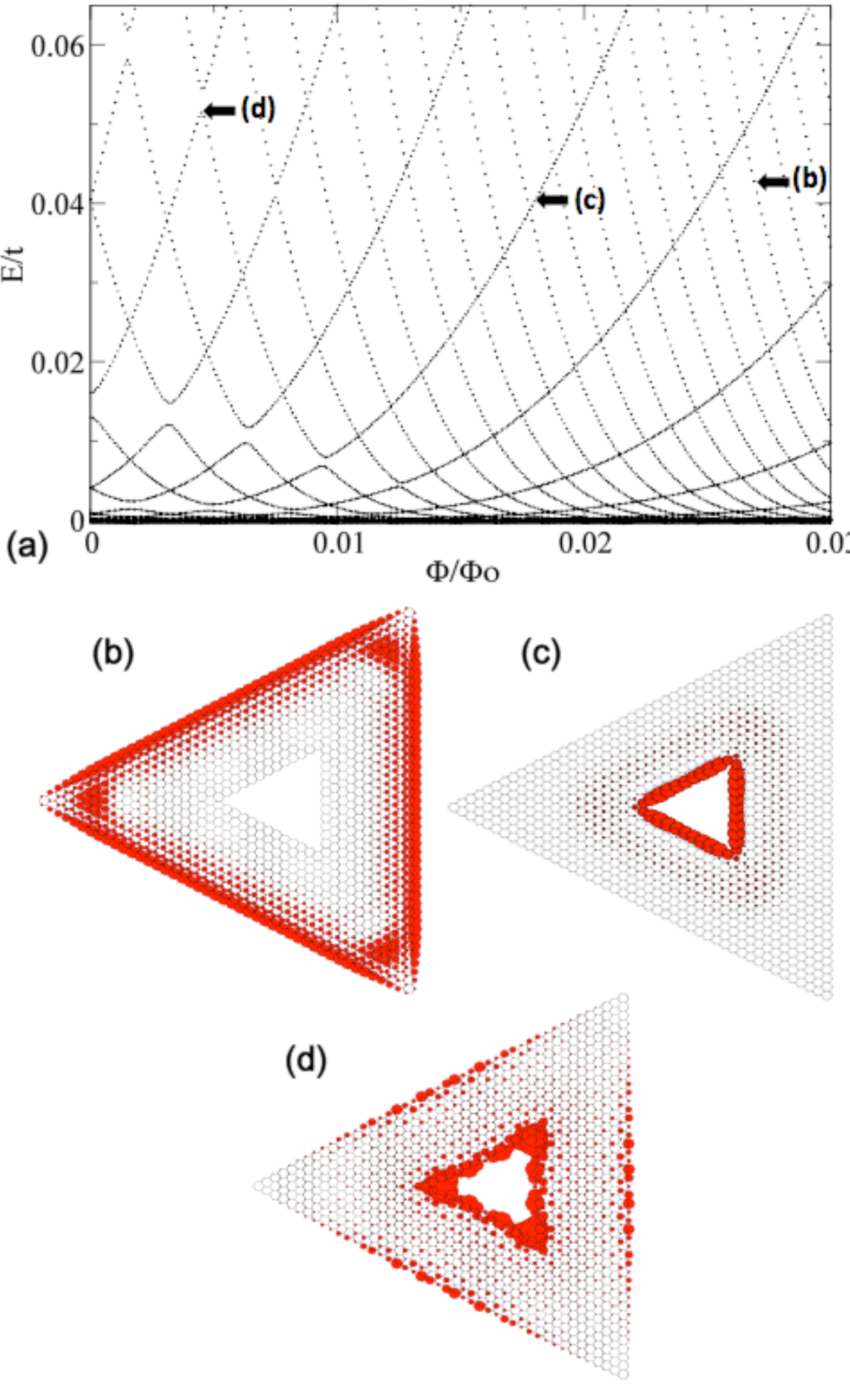}
\caption{ (color online) {\bf (a)} Zoom in the energy scale of the spectrum shown in Fig. 1(d), showing now only the first few low-energy states and their evolution with magnetic flux. {\bf (b-d)} Electronic charge distribution of the selected edge states indicated by the arrows and corresponding letters in the spectrum. Three typical behaviors are clearly defined: {\bf (b)} an state whose energy is reduced with $B$ is an outer edge state. {\bf (c)} an state whose energy is increased with $B$ is an inner edge state. {\bf (d)} at the anticrossing levels the wave functions are distributed between the inner and the outer edges, indicating a coupling between both edges.  The radii of the circles are proportional to the amplitude of the charge density.}
\end{figure}


The arrow (c) in Fig. 2(a)  points to one of those states that go up in energy with $B$, whose wave function is mapped in Fig. 2(c). In agreement with the observation that those states going up in energy appear only when the antidot in considered in the lattice, this is an edge state clearly located at the inner edge of the triangular quantum ring. Its electronic charge density is homogeneously distributed over the innermost lattice sites with decreasing amplitudes when approaching the corners.

This system has a three-fold rotational symmetry. The outermost atoms at the outer edges are all from the same sublattice except for the three atoms located at each corner. The atoms at the innermost edges are all from the same sublattice, including that located at the corners, and this is an important difference between the inner and outer edges in the triangular zigzag quantum ring. The charge density in the innermost edge is also sublattice modulated but with the charge density dominantly on a different sublattice than at the outermost edge.

\subsection{C. Coupling between inner and outer edge states}

An interesting observation emerges from  mapping the charge density of a state situated at an anticrossing, like the state indicated in the spectrum by the arrow (d)  and with the charge density depicted in Fig. 2(d). It can be observed that the wave function has amplitudes concentrated on both the inner and the outer edges of the ring. This indicates  that a coupling between edge states from the inner edge and from the outer edge is taking place.
Consistently to this picture of inner and outer edge states seeing each other and getting coupled, we observe that the higher the magnetic flux, the smaller are the energy gaps between the sub-bands. This can be attributed to the fact that increasing the magnetic flux $\Phi/\Phi_{0}$, the magnetic length $l_B$ of the states gets smaller, reducing the chances of coupling.

For a more quantitative comparison, the width (distance between outer and inner edges) of the triangular ring we are considering (with $N_{out}$=45 and $N_{in}$=12) is 24.1{\AA} and the magnetic length is determined by: $l_B=\sqrt{\hbar /eB}$ = 0.913{\AA}/$\sqrt{\Phi/\Phi_0}$. In this way, for a flux $\Phi/\Phi_0$=0.02, for which there are no energy gaps in the scale observed in the spectrum of Fig. 2(a), we have $l_B$=6.46\AA. Reducing the flux, for example for $\Phi/\Phi_0$=0.01, where energy gaps already start to appear, the magnetic length is $l_B$=9.13\AA. For $\Phi/\Phi_0$=0.005, a region of flux where the energy gaps are more clearly defined, we have $l_B$=12.9\AA, a value corresponding to approximately half the width of the triangular ring, and so compatible with the suggested coupling between outer and inner edge states.

We recall the fact that we are showing typical charge density plots: any other chosen edge state that goes down (or up) in energy has a very similar charge density distribution to that shown in Fig. 2(b) (or 2(c)), while any state at an anticrossing shows wave function concentration on both inner and outer edges, similarly to the distribution observed in Fig. 2(d). It is interesting to notice that the coupling of both edges does not break the sublattice modulation of the charge density at each edge, but the state as a whole is now sublattice mixed.

\section{IV. Widths, edges and corner effects on the quantum rings}

\subsection{A. Widths and sub-band gaps: tuning the inner and outer edge coupling}



\begin{figure}[b]
\includegraphics[width=8.3cm]{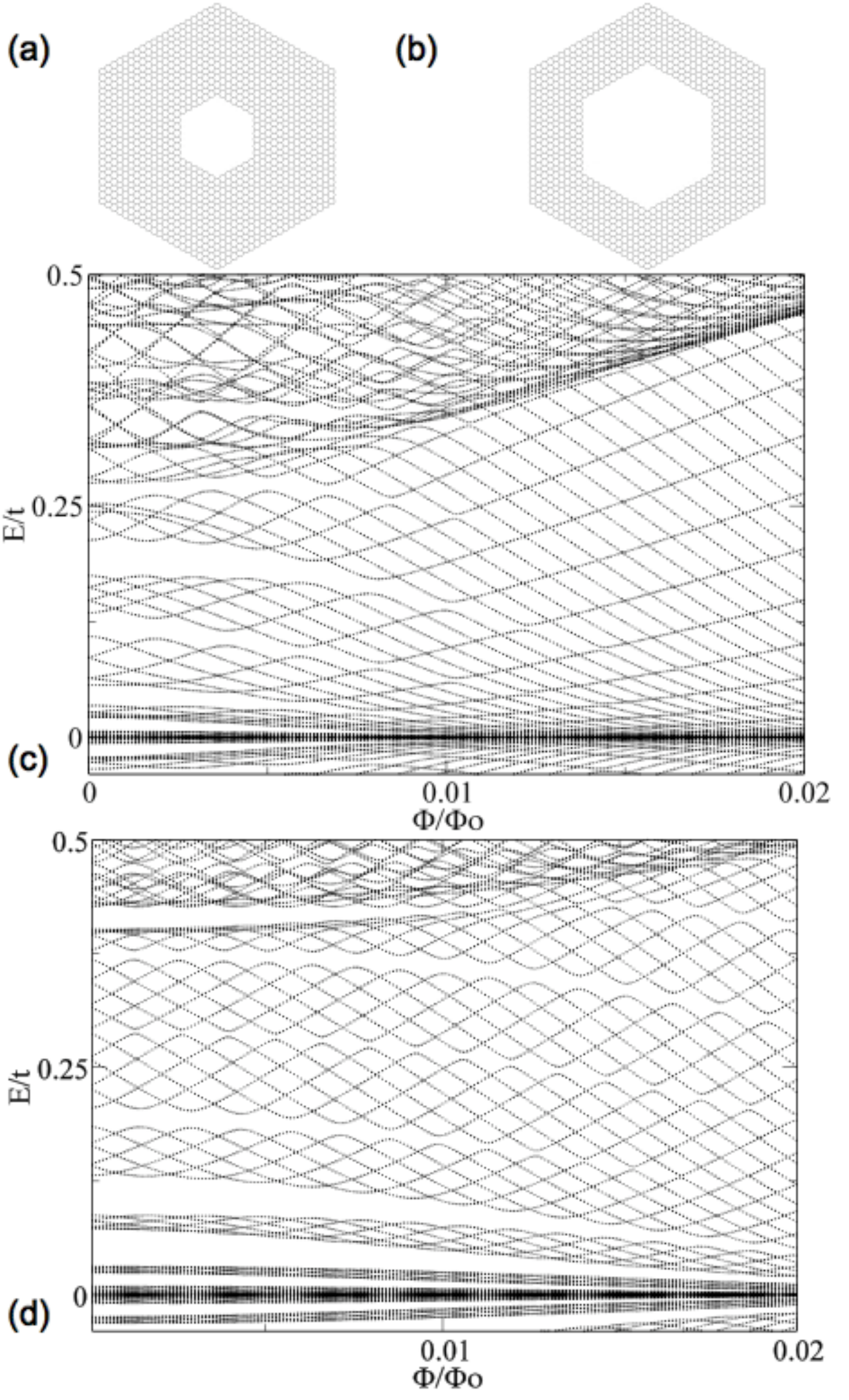}
\vspace{-0.3cm}
\caption{ (color online) {\bf (a)} Hexagonal zigzag quantum ring with $N_{out}=21$ and $N_{in}=7$. {\bf (b)} Thinner hexagonal zigzag quantum ring, with the same $N_{out}=21$, but with $N_{in}=12$. {\bf (c)} Energy spectrum as function of the magnetic flux for the structure in (a). {\bf (d)} Energy spectrum as function of the magnetic flux for the structure in (b).}
\label{Fig3} \end{figure}


We now turn our attention to hexagonal graphene quantum rings, first to compare the energy spectrum of this other geometry with the one from the triangular ring shown previously, and second to show the interesting effects of varying the width of the quantum ring on the formation of the edge states energy sub-bands. In Figs. 3(a) and 3(b) there are representations of two hexagonal quantum rings, with different widths. We once again consider a geometry with all zigzag edge terminations. The total number of atoms in an hexagonal zigzag graphene ring like these is $6N^{2}_{out}-6N^{2}_{in}$, where $N_{out}$ and $N_{in}$ are the number of hexagonal plaquetes along each side of the hexagon and removed hexagon, respectively. Again, the total length of each side of the structure is just given by the number $N_{out}$ or $N_{in}$ times the lattice constant $a$. For the ring in Fig. 3(a) we consider $N_{out}$=21 and $N_{in}$=7, while the ring in Fig. 3(b)  has $N_{out}$=21 and $N_{in}$=12.

The energy spectra as a function of magnetic flux of these two rings are shown in Figs. 3(c) and 3(d), respectively. Comparing these spectra with one of an hexagonal quantum dot (without the antidot in the middle) \cite{peeters}, it is evident that the ring geometry introduces energy sub-bands separated by energy gaps, exactly as in the case of the triangular quantum ring. However, it can now be observed that each band of the hexagonal structures has six energy levels instead of the three levels observed in the triangular structures. We note that this follows the rotational symmetry fold number of the hexagonal ring structure. Here the sublattice of the outer and inner most edges alternates from one sublattice to the other going from one arm of the hexagon to the next \cite{recher,peeters}.

The width of the ring (distance between outer and inner edges) in Fig. 3(a) is 29.8\AA, while the width of the thinner ring in Fig 3(b) is 19.9\AA. When observing the effects of varying the width in the energy spectra, these ring widths can be compared to the magnetic lengths for corresponding magnetic fluxes, as described in the previous section. Corroborating the idea that the coupling between inner and outer edge states is directly related to the appearance of the energy gaps between sub-bands, we clearly see that the thinner quantum ring shows energy gaps in the spectrum until higher values of magnetic fluxes (smaller magnetic lengths).

\subsection{B. Zigzag versus armchairm edges: differences in the quantum ring spectra around the Dirac point}


\begin{figure}[b]
\begin{center}
\includegraphics[width=9.1cm]{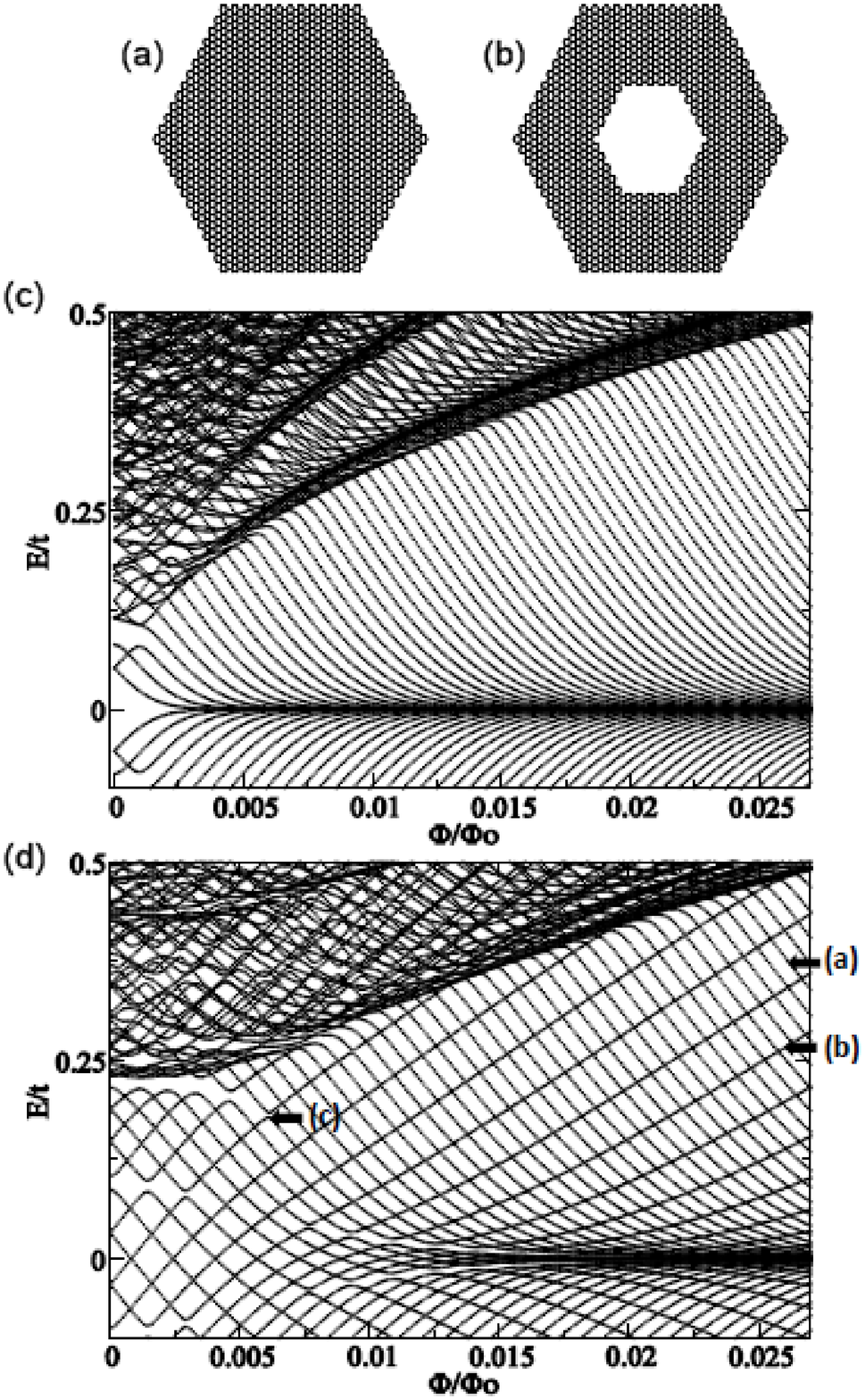}
\vspace{-0.3cm}
\end{center}
\caption{{\bf (a)} Hexagonal graphene quantum dot with armchair edges, for which $N_{out}=13$. {\bf (b)} Hexagonal graphene quantum ring with armchair edges, with $N_{out}=13$ and $N_{in}=6$. {\bf (c)} Energy spectrum for the structure in (a). {\bf (d)} Energy spectrum for the structure in (b), where the arrows indicate states whose charge densities are plotted in Figure 5.}
\label{Fig4} \end{figure}



\begin{figure}[b]
\begin{center}
\includegraphics[width=8.1cm]{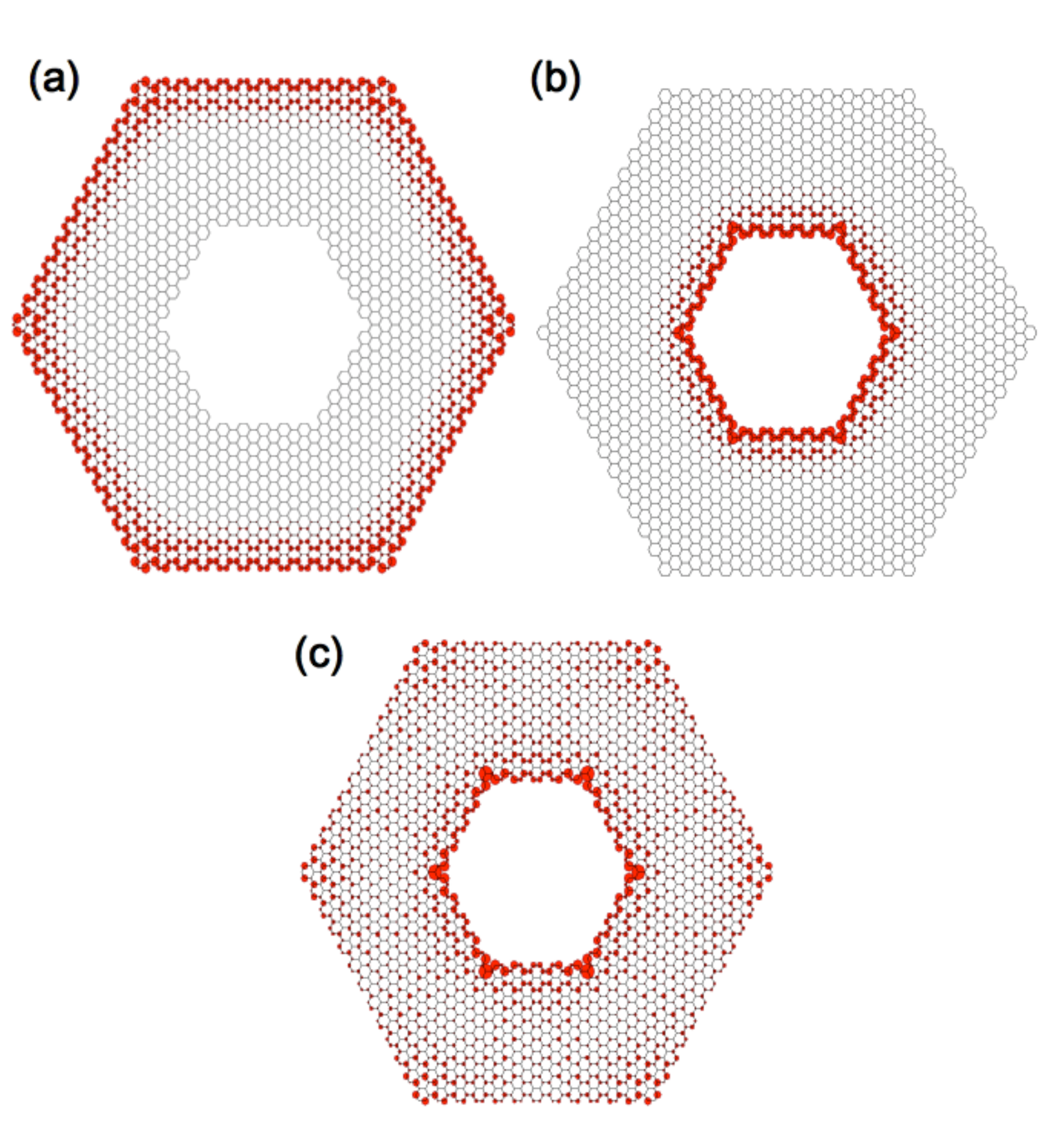}
\vspace{-0.3cm}
\end{center}
\caption{ (color online)  Electronic charge distribution of the selected edge states indicated by the arrows and corresponding letters in Fig. 4(d). {\bf (a)} an outer edge state. {\bf (b)} an inner edge state. {\bf (c)} coupling between the inner and the outer edges, at an anticrossing.}
\label{Fig5} \end{figure}


Having in mind the possible importance of the edge structure on the electronic structure of graphene quantum rings, we now look to hexagonal quantum dot and quantum ring systems with inner and outer armchair edges (Figs. 4(a) and 4(b)). The corresponding electronic structures as a function of magnetic field are shown in Figs. 4(c) and 4(d). The number of hexagonal plaquetes  in each side of the hexagonal dot considered is $N_{out}=13$ (the counting for armchair edge terminations takes in account only the outermost plaquetes), corresponding to a total of 2814 atoms in the nanostructure. For the hexagonal ring, we considered $N_{out}=13$ and $N_{in}=6$, where $N_{in}$ is again the number of hexagonal cells in one side of the hexagon removed.

Recalling that at $B$=0 there are no states associated to armchair edges  near the Dirac point \cite{Wakabayashi}, the central part  (around $E$=0) of the quantum ring spectrum here is completely different than in the case of zigzag edges (compare with Fig. 3). There are still edge states sub-bands defined, each one containing six energy levels, however the difference is that there is now a wide sub-band around the Dirac point. An interesting observation is the interchange between electron-like and hole-like states in this region, as a function of magnetic field. The huge difference in the electronic dispersion should be reflected in the related transport properties.

Similarly to the zigzag case, a clear and strong localization of the charge density at the inner and outer edge occurs in the decoupled edges limit (high magnetic field), as observed in the examples of Fig. 5(a) and Fig. 5(b), corresponding to the states pointed by the arrows (a) and (b) in  the ring spectrum (Fig. 4(d)). Around an anticrossing, as for the state pointed by arrow (c), the charge density, observed in Fig. 5(c) is spread out on the two edges, indicating the edge coupling. As a difference between zigzag and armchair cases edges, we see that an armchair termination leads to a sublattice admixture of the charge density, different from the case of the zigzag edges, were there is a sublattice modulation \cite{Brey1}.

\subsection{C. Asymmetries introduced by the corners in diamond rings}

Next we consider a rhombus-shaped (diamond) graphene quantum ring which has only zigzag edges. This ring is interesting because of its two-fold rotational symmetry and because the upside outer edges  of the diamond (in the perspective of the pictures in Figs. 6(b-d)) are from one sublattice meanwhile the downside outer edges are from the other sublattice. However, similarly to the triangular quantum ring, the up and down corners belong to a different sublattice of the neighboring edges. The same sublattice effects occur at the inner edges. The number of atoms for this kind of ring is given by $2(N^{2}_{out}-N^{2}_{in})+4(N_{out}-N_{in})-2$, within our definition. The energy spectrum of a diamond-like ring defined by $N_{out}=32$ and $N_{in}=10$ (1934 atoms) is plotted  in Fig. 4(a). One can clearly observe the evolution of the two level bands, as expected from the symmetry of the structure.


\begin{figure}[t]
\begin{center}
\includegraphics[width=8.5cm]{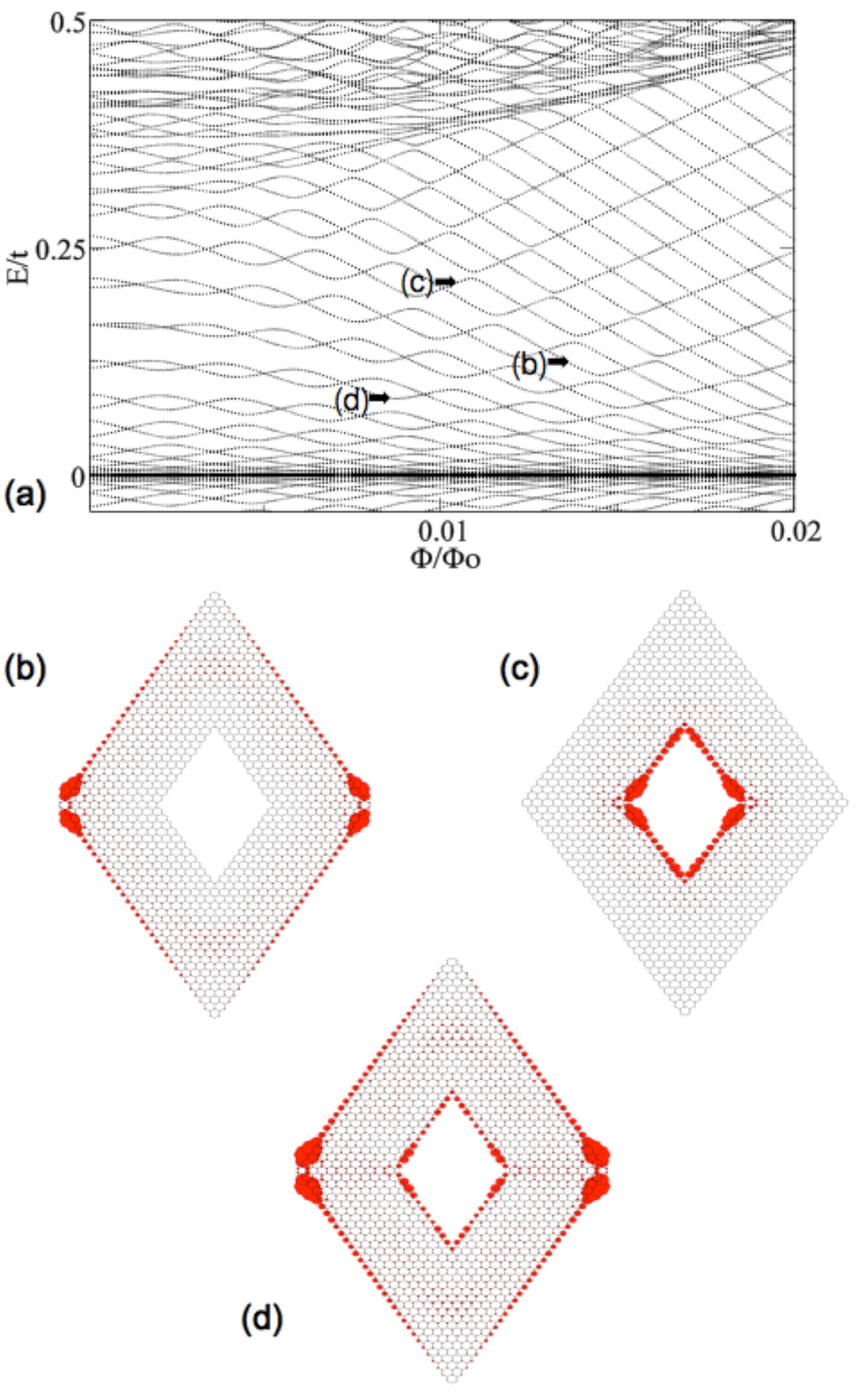}
\vspace{-0.3cm}
\end{center}
\caption{ (color online) {\bf (a)} Energy spectrum as function of the magnetic flux for a rhombus-shaped quantum ring with zigzag outer and inner edges, containing $N_{out}=32$ and $N_{in}=10$ hexagonal plaquetes in each side length. Electronic charge densities: {\bf (b)} going down state marked with the (b) arrow in the spectrum, {\bf (c)} the going up state marked with the (c) arrow and {\bf (d)} for an anticrossing state marked with the (d) arrow. The radii of the circles denote the magnitude of the charge density.}
\label{Fig6} \end{figure}


All the rings investigated in the present work with zigzag edges show similar energy spectra: at the low field limit sub-bands of energy states are formed, with the number of levels in each sub-band given by the ring symmetry and well defined outer and inner edge states at higher magnetic fields. Nevertheless, a closer look at the electronic charge density associated to different up and down going states, as well as at anticrossings, for these diamond-like rings reveals a further ingredient in the effect of edges on the electronic properties of quantum dots and rings in graphene, namely the edge junctions at the corners. In a diamond-like ring the junctions between the zigzag edges define single armchair-like units at the left and right (inner and outer) edges (Figs. 6 (b)-(d)), while the upper and lower corners remain zigzag like. One can see, in the sequences of electronic charge distributions in Fig. 6, the high density around the armchair like corners, independently from being a state at the outer edge, Fig. 6(b), inner edge, Fig. 6(c), or even at an anticrossing, Fig. 6(d).

This situation calls the attention to the possible role of the edge junctions on the localization of the electronic charge in graphene nanostructures, i.e., the localization of the electronic charge at a rough interface may depend on the symmetries at the corners that define the edge landscape.

\subsection{D. Round rings - effects of irregular edges}

We then analyze the cases of a round graphene dot and ring. Here the edges of the structures are irregular and were defined in a way to best approach circular geometries for outer and inner edges, taking care not to leave edge atoms with only one nearest-neighbor  \cite{akola}, as observed in Figs. 7(a) and 7(b).
Figures 7(c) and 7(d) show the corresponding energy-magnetic flux spectra for these two structures. For the round dot, the number of atoms that has been taken is 2283, defining a radius of $\approx47.1$ \AA. For the round ring, the external radius is $\approx 47.1$ \AA, and the internal radius is $\approx 7.3$ \AA, containing a total of 2226 atoms.


\begin{figure}[b]
\begin{center}
\includegraphics[width=8.5cm]{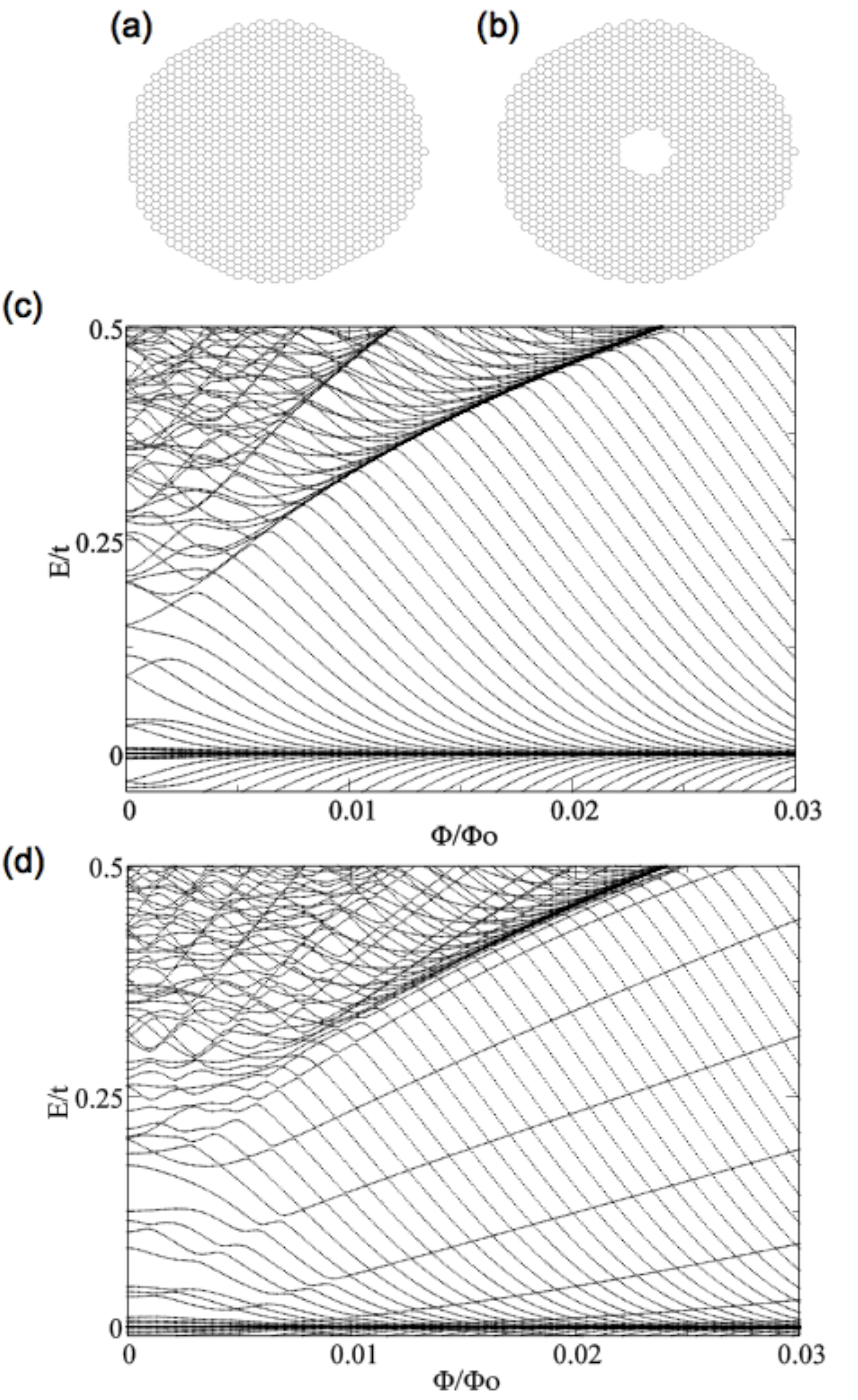}
\vspace{-0.3cm}
\end{center}
\caption{ {\bf (a)} Round graphene quantum dot with 2283 atoms. {\bf (b)} Round graphene quantum ring with 2226 atoms. {\bf (c)} Energy spectrum for the structure in (a). {\bf (d)} Energy spectrum for the structure in (b).}
\label{Fig7} \end{figure}


One can perceive from these spectra that, despite the irregularities of the edges, the main effects observed from the previous geometries are robust and keep present here. Comparing the spectra for the circular ring (Fig. 7(d)) with the one for the circular dot (Fig. 7(c)), it is again clear that the circular antidot introduces inner edge states whose energies are increased with increasing magnetic flux. In the low flux limit of the ring spectra, anticrossing levels are again observed in the edge states region, indicating the coupling of inner and outer edge states, exactly as observed and described for the structures with well defined edge structures (zigzag or armchair). The main difference is that, as for this geometry there is no rotational symmetry, we do not observe the formation of edge-states sub-bands with well defined number of energy levels.

\section{V. Conclusions}

The present paper focus on the single particle electronic properties of finite graphene structures.
The behavior of edge states in graphene rings is investigated, through the numerical calculation of the electronic energy spectra of these rings as a function of a perpendicular magnetic field and the mapping of charge density distributions. Several similar patterns may be found among quantum rings with different symmetries (triangular, hexagonal, diamond shapes), including the formation of sub-bands of edge states energy levels, separated by energy gaps (anticrossings). The choice of quantum rings revealed a strategic one because of the clear relation between the symmetry of the structure and the number of levels in the edge states sub-bands. Furthermore, the edge states levels within the sub-bands can be perfectly associated to either inner or outer edges, as well as the ``bulk" region of the structure (coupling between edges), therefore enabling a good framework for studying the influence of the edges on the electronic structure and charge distribution. If edge terminations (zigzag or armchair) show to play an important role on the electronic properties, specially for the states around the Dirac point, the junction of the edges (corners) can also be crucial for charge density localization patterns.


\section{Acknowledgments}

DABA acknowledges support from CAPES, ALCP acknowledges support from FAPESP. PAS received partial support from CNPq.


\end{document}